\documentstyle[12pt]{article}
\input psfig.tex
\begin{document}
\begin{center}{\large\bf Exclusive Hadronic
Processes
and Color Transparency}
\end{center}

\bigskip
\centerline{\large\bf Pankaj Jain$^a$ and John P. Ralston$^b$}

\bigskip
\begin{center}{$^a$Physics Department, I.I.T. Kanpur, Kanpur - 208016\\
$^b$Department of Physics and Astronomy, University of Kansas,\\
Lawrence, KS 66045 USA}
\end{center}

{\bf Abstract:}
We review the current status of high energy exclusive processes 
and color transparency.

\section{Introduction}
It is known that at {\it asymptotically large momentum transfer}
certain exclusive hadronic reactions are calculable within the framework of
perturbative QCD (pQCD) due to asymptotic freedom.  However the
applicability of pQCD to exclusive processes remains controversial at
the largest momentum transfers probed in the laboratory.  The
quark-counting scaling laws of Brodsky and Farrar \cite{GRF} tend to
agree remarkably well with data.  This apparently indicates that a
finite, minimal number of quarks are being probed.  However, the
helicity conservation selection rules of Lepage and Brodsky tend not
to agree with data \cite{GRF,GPL,CZ}.  This indicates that the
Brodsky-Lepage factorization scheme fails, independent of further
details.  Furthermore the formalism fails to correctly predict the
magnitude of the proton electromagnetic form factor \cite{Isg,Rad,SS}. 
This suggests that the basic assumption in this formalism, namely that
these processes are dominated by short distance, may not be
true at laboratory energies.  As a consequence, the agreement of the scaling
laws with data becomes rather mysterious.

Compared to exclusive processes in free space, it has been shown
\cite{JB,RP90,PBJ} that the corresponding processes in a nuclear medium will 
be
theoretically cleaner. Large quark separations will tend
not to propagate in the strongly interacting nuclear medium.  Configurations 
of
small quark separations, on the other hand, which coincide with the
perturbatively calculable region, will propagate with small attenuation.  This
phenomenon, called nuclear filtering\cite{JB,RP90,PBJ}, is the
complement of the
idea called color transparency \cite{brodMuell}. 
In its original rendition,
color transparency \cite{brodMuell} was based on having large momentum
transfer $Q^{2}$ select short distance, then free to propagate easily through 
a
passive nuclear probe. Nuclear filtering uses the nuclear medium in an active
way toward the same purpose, and may be more efffective.

\section{Exclusive Processes in Free Space}

Let us briefly review the framework for calculation of exclusive
hadronic processes within pQCD. We take the pion electromagnetic form
factor as an example.  The short-distance formalism crucially depends
on the assumption that the process can be factorized in a
perturbatively calculable hard scattering piece and the soft
distribution amplitude.  Given this assumption, the pion
electromagnetic form factor \cite{GPL,FJ,ER} at momentum transfer $q^2
= -Q^2$ can be written as
\begin{equation}
F_\pi(Q^2) = \int dx_1 dx_2 \phi(x_2,Q) H(x_1,x_2,Q) \phi(x_1,Q),
\end{equation}
where $\phi(x,Q) $ are the distribution amplitudes which can
be expressed in terms of the pion wave function $\psi(x,\vec k_T)$ as
\begin{equation}
 \phi(x,Q) = \int^Q d^2k_T\psi(x,\vec k_T) .
\end{equation}
Here $x$ is the longitudinal momentum fraction and $\vec k_T$ the
transverse momentum carried by the quark; $\psi$ is the light-cone 
Bethe-Salpeter amplitude.  The factorization is
a good approximation provided the external photon momentum $Q^2$ is much 
larger
than all other physical scales.  The hard scattering is then evaluated 
with on-shell quarks carrying neglible $k_{T}$.

The formalism predicts that the cross section for exclusive processes
$d\sigma/dt$, where $t$ is the momentum transfer squared, scales like
$1/t^{n-2}$ up to logs, where $n$ is the total number of elementary
partons participating in the process.  The underlying reason for the
power law is scale invariance of the fundamental theory.  Further
logarithmic dependence is given by QCD scaling violations.  In making
these assertions one asumes that $t$ is asymptotically large.  The
dominant contribution to this scattering arises from the valence
quark, since every additional parton leads to an additional
suppression factor of $1/t$.  Physically the scattering probes the
short distance part of the hadron wave function.  Dominance by the
short distance wave functions leads to several predictions such as
helicity conservation, color transparency \cite{brodMuell,PBJ} etc.

The successes and failures of this scheme are well known.  Calculation
of electromagnetic form factors using this factorization scheme has
been criticised by several authors \cite{Isg,Rad}.  The basic problem
is that the momentum scales of the exchanged gluons tend to become
rather small, and the applicability of pQCD becomes doubtful.  The
normalization of form factors is largely unknown; use of asymptotic
distribution amplitudes tends to give small normalizations compared to
data.  Form factor magnitudes can be enhanced by use of model
distribution amplitudes \cite{CZ,KS} which peak closer to the
end-points, namely $x\rightarrow 0,1$, which then exacerbates the
problem of small internal momentum transfers.

To investigate this problem, there exists an alternate factorization
which does not neglect the $k_T$ dependence of the hard scattering. 
The method, which we call ``impact parameter factorization'', was
first used by Botts and Sterman \cite{BS} to deal with the Landshoff
pinch regions of proton-proton scattering.  These regions lie outside
the assumptions of the quark-counting formalism, and cannot be
described by its factorization scheme.  Impact parameter factorization
was first applied to electron-beam ($\gamma^{*}$ initiated)
experiments in Ref.  \cite{RP90}, in order to accomodate color
transparency and nuclear filtering.  Li and Sterman \cite{LS}
developed the method for free-space $\gamma^{*}$ initiated experiments
such as the proton's electromagnetic form factor.  A consistent
feature of this formalism involves attention to the transverse 
spatial coordinates of quarks and attendant Sudakov effects.

For the case of pion form factor
\cite{LS}
the starting point is,
\begin{equation}
F_\pi(Q^2) = \int dx_1 dx_2 d\vec k_{T1} d\vec k_{T2} \psi^*(x_2,\vec
k_{T2},P_2)
H(x_1,x_2,Q^2,\vec k_{T1},\vec k_{T2})
\psi(x_1,\vec k_{T1},P_1),
\end{equation}
where it is again assumed that the process is factorizable into
hard scattering and soft hadronic wave functions $\psi(x,\vec k_T,P)$. The
calculation is simplified by dropping the $k_T$ dependence in the
quark propagators in hard scattering kernel $H$,
in which case only the combination $\vec k_{T1} + \vec k_{T2}$
appears in the calculation.
The authors \cite{LS} work in configuration space where this can be written
as
\begin{equation}
F_\pi(Q^2) = \int dx_1 dx_2 {d^2\vec b\over (2\pi)^2} {\cal P}(x_2,b,P_2,\mu)
\tilde H(x_1,x_2,Q^2,\vec b,\mu) {\cal P}(x_1,b,P_1,\mu),
\end{equation}
where ${\cal P}(x,b,P,\mu)$ and $\tilde H(x_1,x_2,Q^2,\vec b,\mu)$ are the
Fourier transforms of the wave function and hard scattering
respectively;
$\vec b$ is conjugate to $\vec k_{T1} + \vec k_{T2}$, $\mu$ is the
renormalization
scale and $P_1$, $P_2$ are the initial and final momenta of the pion.

Sudakov form factors are obtained by summing the leading and next to
leading logarithms using renormalization group (RG) techniques.  The
wave function at small $b$ is often approximated by the distribution
amplitude $\phi(x,1/b)$. 
\begin{figure} [t,b]
\hbox{\hspace{6em}
 \hbox{\psfig{figure=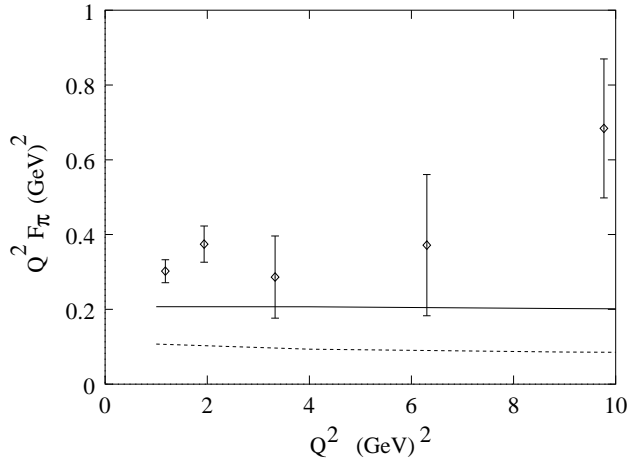,height=6cm}}}
  \caption {\em The pion form factor $F_\pi (Q^2)$
using the asymptotic (dotted line) and the CZ
    (solid line) distribution amplitudes.
    The experimental data with error bars from Ref. [16] are
also shown.}
  \label{pion1}
\end{figure}

The resulting form factor using asymptotic as well as CZ \cite{CZ}
distribution amplitudes is shown in fig.  \ref{pion1}.  A remarkable
fact is that the correct asymptotic $Q^2$ behavior is seen beyond the
scale of about $Q=1$ GeV, irrespective of the choice of wave function. 
In contrast to the Brodsky-Lepage factorization, the $k_T$ dependence
of the hard scattering is not neglected, and hence this $Q^2$
dependence does not follow trivially.  Instead, the $Q^{2}$ dependence
is a detailed dynamical prediction of the theory, and depends
on the relative size of intrinsic $k_T^2$ and $x_1 x_2 Q^2$.  The
$Q^2$ dependence of the prediction is comparatively robust, since it
is weakly dependent on the details of the distribution amplitude.

We note that the normalization of the theoretical result falls below
the experimental data for both choices of distribution amplitude.  It
is good to keep in mind that the distribution amplitudes are not known
with much exactitude, so perhaps the models might be improved.  Indeed
the theoretical normalization of the form factor is comparatively
murky, because it strongly depends on such model-dependent details. 
Moreover, the large difference between theory and experiment at high
momenta should be interpreted with caution, since\cite{SS} there may
be large systematic errors in the experimental extraction of the form
factor which are not shown in the figure.  Further theoretical issues
in this extraction have also been raised in Ref.  \cite{CM}.

Furthermore, the leading order pQCD amplitudes calculated may not give
a very reliable estimate of the normalization.  Li and Sterman argue
that roughly 50\% of the contribution can be regarded as perturbative,
since it is obtained from the region where $\alpha_s/\pi < 0.7$.  It
may be that higher order contributions in $\alpha_s$ are not
negligible, and the leading order predictions for the normalization of
the form factor cannot be regarded as accurate.

We are left with the following interesting situation: Although the
basic Brodsky-Lepage factorization survives at asymptotic $Q^{2}$, the
method is sensitive to end-point singularities, and one may
need to go to higher orders in $\alpha_s$ in order to obtain an
accurate prediction for the form factor normalizations.  Meanwhile the
predicted $Q^2$ dependence of the impact parameter factorization
appears to be quite robust, and less strongly dependent on the
theoretical uncertainties such as the choice of distribution
amplitude.

\subsection{The Proton Electromagnetic Form Factor}

\begin{figure} [t,b]
\hbox{\hspace{6em}
 \hbox{\psfig{figure=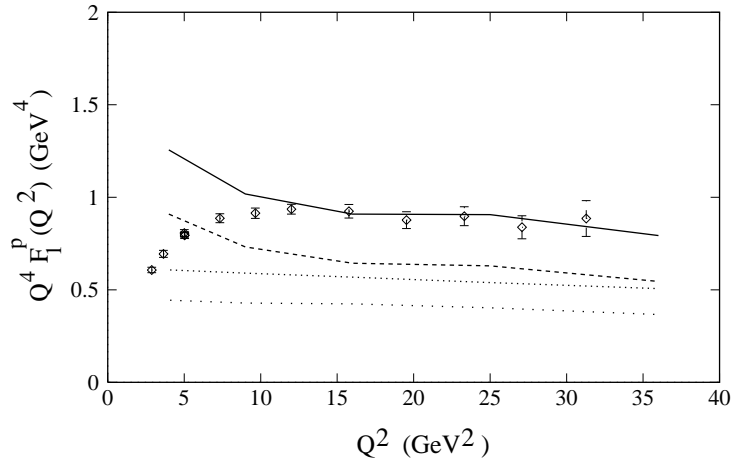,height=6cm}}}
  \caption {\em Dependence of proton form factor $Q^4F_1^p$ on $Q^2$ using the
    KS distribution amplitude ($c=1.14$, solid line; $c=1$, dense-dot line)
     and the CZ distribution amplitude ($c=1.14$, dashed line;
               $c=1$, dotted line).
    The experimental data with error bars, taken from Ref. [20],
 are also shown.}
  \label{proton1}
\end{figure}

The improved impact parameter factorization has also been applied to
the proton Dirac form factor $F^p_1(Q^2)$\cite{LS1}.  The calculation
is considerably more complicated compared to the pion due to the
presence of three valence quarks.  Here also it is necessary to use
distribution amplitudes which peak close to the end points if one
wishes to fit the experimental normalization of the form factor. 
The results \cite{LS1,BL} of the calculation using KS \cite{KS} and CZ
\cite{CZ} distribution amplitudes and $c=1$ and 1.14 are shown in fig. 
\ref{proton1}, where $c$ is a parameter which determines the long and
short distance factorization scale.

The natural agreement of $Q^{2}$ dependence of the pQCD calculations
should be contrasted to data fits obtained using soft overlap models
\cite{Rady,Diehl}.  In such models the $Q^2$ dependence depends on the
details of the model wave function.  Soft overlap model predictions at
high momentum have a tendency to fall more strongly than experimental
data.

\subsection{Hadron-Hadron Exclusive Processes}

To a rough approximation, the empirically fit power-laws for
hadron-hadron exclusive processes tend to agree well with the
quark-counting scaling law.  However when data is examined in detail,
one finds contradictions, such as the violation of helicity
conservation selection rules and oscillations
\cite{pire82,helicity81,pip} around the overall power law momentum
dependence of $d\sigma/dt$.  It is not commonly appreciated that these
signals of processes beyond the quark-counting model are observed for
almost every process tested so far.

\begin{center}
\begin{figure}
\psfig{file=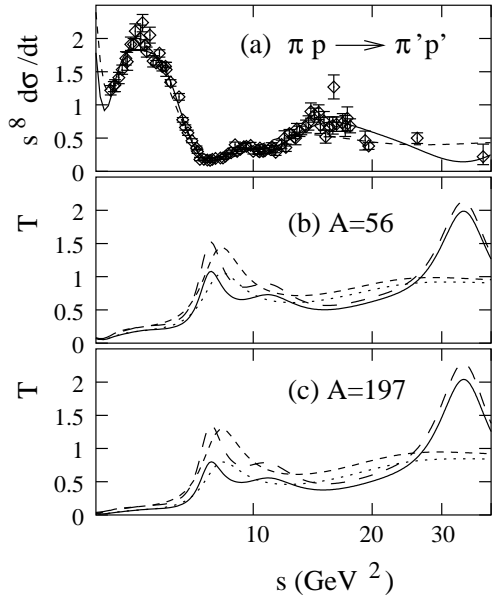}
\caption{\em (a) The free space $\pi p$ $90^o$
cross section $s^8 d\sigma/d|t|$ ($10^8$ GeV$^{16}$ $\mu$b/GeV$^2$)
using three component model [25] (solid curve) and a two
component model (dashed).  (b,c) Calculated color transparency ratio
for $A=56,197$ using nuclear filtering, in the three component model
with $k=10$ (solid), $k=5$ (long dashed) and the two component model
with $k=10$ (dotted) and $k=5$ (short dashed).}
\label{pip}
\end{figure}
\end{center}

\subsection{Free Space Data: $\pi$-P to $\pi$-P at Fixed Angle}

In Fig.  \ref{pip}a we show results $d\sigma/dt$ for $\pi p\rightarrow
\pi' p'$ scattering at $90^{o}$\cite{pip}.  An overall momentum
dependence $|t|^{-8}$ has been factored out of the cross section
\cite{owen}.  The free space cross section is fit in terms of
interfering short and long distance amplitudes using a three component
(solid curve) and two component model (dashed curve).  One amplitude
is obtained from the independent scattering diagrams \cite{Landshoff},
where the quarks scatter at large transverse separations.  Leading
logarithmic summation calculable in pQCD and related by analyticity
to Sudakov effects generates the oscillations \cite{pire82,BS}.  In
the three component model we also include a subleading contribution:
details are given in Ref. \cite{pip}.  We note that the oscillations
around the overall power dependence are not a small effect.  The model
incorporating the independent scattering diagrams fits the free space
data very well with $\chi^2/{\rm degree\ of\ freedom} = 1.97$.  The
Brodsky-Lepage short distance model gives $\chi^2/{\rm degree\ of\
freedom} = 99$ and is objectively ruled out.

\subsection{Free Space Data: $\gamma$-P to $\pi$-N at Fixed Angle}

\begin{center}
\begin{figure}
\psfig{file=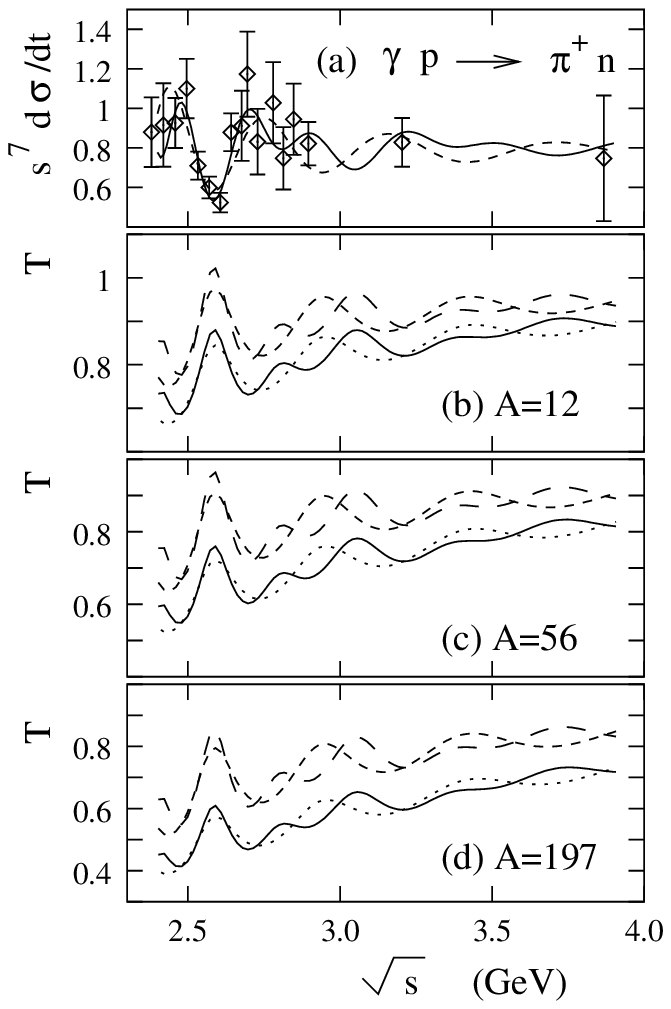}
\caption{\em (a) The free space $\gamma p
\rightarrow \pi^+ n$ $90^o$ cross section $s^7d\sigma/d|t|$ ($10^7$
GeV$^{14}$ nb/GeV$^2$) using three component model (solid curve) and a
two component model (dashed).  (b,c,d) Calculated color transparency
ratio for $A=12,56,197$ using nuclear filtering, in three component
model with $k=10$ (solid), $k=5$ (long dashed) , and in two component
model with $k=10$ (dotted) and $k=5$ (short dashed).
}
\label{gammap}
\end{figure}
\end{center}

The experimental data \cite{anderson} for $\gamma p\rightarrow \pi^+n$
also shows fluctuations around the overall power behavior as seen in
Fig.  \ref{gammap}a.  This is quite interesting, since in this case
the Landshoff pinch amplitudes have been theoretically shown to be
subleading at large $Q^2$ \cite{FSZ}.  However at medium $Q^2$ even
the subleading amplitudes can give significant contributions.  We
again model the fluctuations in terms of interfering soft and hard
contributions.  The free space fit is shown in Fig.  \ref{gammap}a.

\section{Nuclear Targets and Color Transparency}

Color transparency is a natural prediction of pQCD. If exclusive
processes at large momentum transfer are actually dominated by short
distance, then interaction of hard-struck hadrons with other hadrons
is predicted to be small.  The hadron interaction cross section $\sigma
\propto b^2$, where $b$ is the transverse separation between quarks in
a color singlet state, vanishes as $b \sim 1/Q \rightarrow 0$. 
Experimentally color transparency is measured by observing quasi
exclusive reactions on nuclear targets such as $eA \rightarrow
e'p(A-1)$ and comparing it with the corresponding free space process
$ep\rightarrow e'p'$.  The experimental results are reported in terms
of transparency ratio $T$, defined as $$ T = {d\sigma_{\rm
nuclear}\over A d\sigma_{\rm free\ space}}\ .
$$

At asymptotically large $Q^{2}$, one has asymptotically short
distance, and color transparency ratios are unity.  This is actually
as much as one can say with quark-counting factorization.  To say more
one must incorporate information about the transverse, or impact
parameter separation, hence the use of ``impact parameter
factorization''\cite{RP90}.

There are two ways, then, to get to short distance: asymptotically
large $Q^{2}$, which we have seen above is not a realistic feature of
the laboratory, or by {\it nuclear filtering}.  The nucleus
essentially acts as a transverse-separation filter \cite{RP90} which
preferentially attentuates the soft amplitudes.  It follows that the
nucleus is a cleaner medium to study exclusive processes than free
space: which is quite a surprising prediction.
 
Several experiments indicate that color transparency \cite{brodMuell}
and nuclear filtering \cite {JB,RP90,JR93,BT88} have been observed at
large nuclear number $A$.  The first color transparency experiment of
Carroll {\it et al} \cite {Car} convincingly showed that interference
effects in proton-proton scattering were filtered away in nuclear
targets.  Attenuation of the Landshoff-induced oscillations, seen in
free-space \cite{pire82}, in the nucleus reproduces the oscillating
transparency \cite{JB}.  In contrast, other models based on a
classically expanding cross section \cite{FFSL} do not fit the data. 
Extraction of the attenuation cross section in nuclear targets show
values significantly below the Glauber theory values \cite {PBJ}.  The
FNAL E-665 experiment \cite {Fang} also proved consistent with
filtering effects \cite {KNN93}.

Electron beam experiments remain controversial, with few signals of
interesting $Q^{2}$ dependence \cite{Mak}.  A basic feature of
$\gamma^*$-initiated reactions is that most events are knocked out
from the back side of the nucleus.  The resolving power of such
experiments to measure the size of propagating states is rather
modest, unless one has very high experimetal precision. The $A$
dependence is a particularly useful tool \cite {PBJ} to measure
effective attenuation cross sections.  O'Neill {\it et al} \cite
{Neill} showed that effective attenuation cross sections extracted
from $ A (e, e'p)$ SLAC data were smaller than Glauber theory
calculations by a statistically significant amount.  However, the
precision of the data \cite{Mak} was insufficient to establish a large
effect, and model dependence in the choice of the normalization of
hard scattering is another complication.  Reports on new $(e, e'p)$
beam experiments from CEBAF are expected shortly.

\medskip
\begin{figure} [t,b] \hbox{\hspace{6em}
\hbox{\psfig{figure=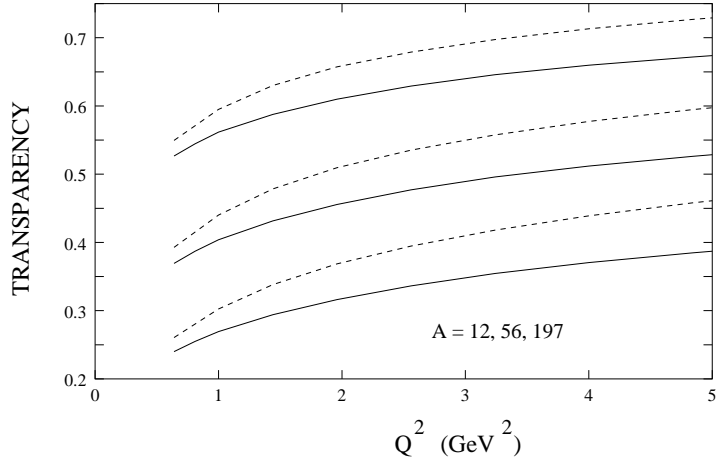,height=6cm}}}
\caption{\em The calculated pion
transparency ratio for different nuclei as a function of $Q^2$. 
The solid and dashed curves use the CZ and asymptotic distribution 
amplitudes respectively and correspond to $A=12$, 56 and 197 from top
to bottom. 
More details are given in Ref. [38].
} 
\label{pionQ} \end{figure}

\subsection{New Color Transparency Calculations}

We have recently calculated color transparency ratio for several
processes which include $\pi A\rightarrow \pi' p (A-1)$, $\gamma
A\rightarrow \pi N (A-1)$, $eA\rightarrow e'p(A-1)$ and $eA\rightarrow
e'\pi A$.  For the case of $eA\rightarrow e'p(A-1)$ and $eA\rightarrow
e'\pi A$ we did detailed impact-parameter calculations using leading
order pQCD including the Sudakov effects.  The more complicated
processes $\pi A\rightarrow \pi' p (A-1)$, $\gamma A\rightarrow \pi N
(A-1)$ were modelled in terms of interfering long and short distance
amplitudes following the earlier work on proton proton quasi elastic
scattering in nuclear medium \cite{JB,JR93}.  The magnitude of short
and long distance amplitudes were obtained phenomenologically by
making a fit to the corresponding free space process.

In Fig.  \ref{pionQ} we show the $Q^2$ dependence of the transparency
ratio for electroproduction of pions using the CZ and asymptotic
distribution amplitudes \cite{KSJR}.  The scale of $Q^2$ ranging up to
5 GeV$^2$ may benefit from explanation.  At the exclusive production
point, the relativistic boost factor of a pion is given by $\gamma=
Q^2/(2 m^2_{\pi}) \sim 25 (Q^2/{\rm GeV}^2)$.  Since even a 1 GeV pion
is highly relativistic, we may suppose that the perturbative
calculations may well apply in the comparatively small $Q^2$ regime. 
These calculations show a rather striking rise with $Q^2$ of the
transparency ratio, which should be easily observable experimentally. 
The {\it fact} of a rise does not depend much on the distribution
amplitude, but the {\it slope} of the rise does: we discuss the
reasons shortly when we review the proton.  For these calculations
we used $\Lambda_{QCD} = 200$ MeV. We adjusted the value of the
parameter $k$, defined by $\sigma_{\rm attenuation} = k b^2$, so that
the predicted results for proton (discussed later) are in agreement
with the SLAC data \cite{Mak,Neill}.  This selects the value of $k$ to
be approximately equal to 10.

\begin{figure} [t,b] \hbox{\hspace{6em}
\hbox{\psfig{figure=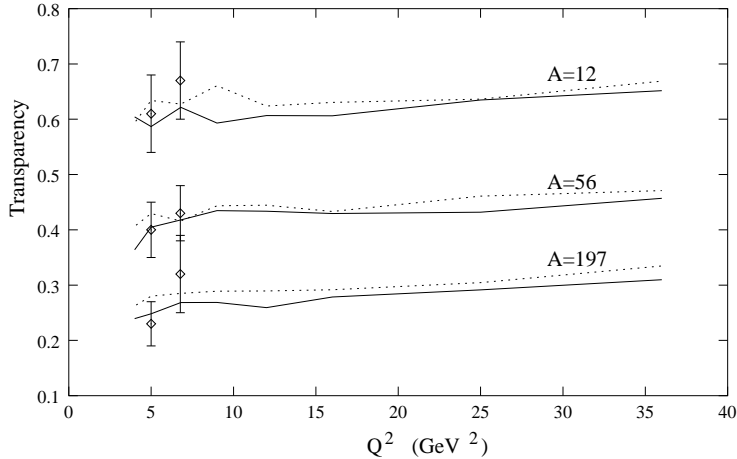,height=6cm}}}
 \caption{\em Predicted
transparency ratio [38] for the proton for different nuclei using the
KS end-point dominated model for distribution amplitude.  The
experimental
points are
taken from Ref. [26,27]. The solid curves are calculated with $k=10$ and the
dashed
curves with $k=9$.} \label{protonQ}
\end{figure}

\begin{figure} [t,b] \hbox{\hspace{6em}
\hbox{\psfig{figure=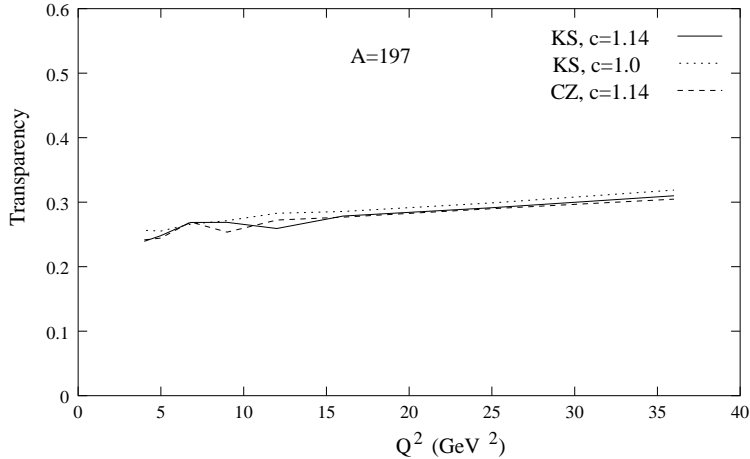,height=6cm}}}
\caption{\em The sensitivity of 
the
calculated transparency ratio to different proton distribution amplitudes and
the factorization scale parameter $c$.
The solid, dotted and dashed curves correspond to
the KS wavefunction with $c=1.14$, KS distribution amplitude with $c=1.0$ and
 the CZ distribution amplitude with $c=1.14$ respectively.
All calculations use $A=197$.  More details are given in Ref. [38]}.
\label{sensitivity}
\end{figure}

Results for the $Q^2$ dependence of the proton transparency ratio
\cite{KSJR} from various models are shown in Figs.  (\ref{protonQ},
\ref{sensitivity}).  A standard model for the distribution amplitude,
the KS model, was used to generate Fig.  \ref{protonQ}.  One sees that
the calculation with the KS model has a rather flat $Q^2$ dependence. 
At first this result was surprising, assuming short-distance dominance
and the expectation of a rapidly increasing function of $Q^2$, but in
retrospect the result appears quite natural.  As in the earlier
discussion of the pion, these results depend on the distribution
amplitude model, which can be categorized into two types.  The KS
model is an {\it end-point} dominated distribution amplitude, which is
known to produce its dominant contributions from long-distance
components of the quark wave functions.  For this reason use of the KS
wave function in the free-space form factor has led to many questions
of theoretical consistency, mentioned earlier.  Precisely the same
lack of a dominant short-distance contribution is responsible
for the calculated flat dependence on $Q^2$.  Turning to Fig. 
\ref{sensitivity}, which compares the CZ and KS models, both of which
are end-point dominated, one sees nearly identical flat $Q^2$ behavior. 
This indicates that the details of the model do not matter so long as
they are end-point models.  The figure also shows the dependence on the
factorization scale parameter $c$.  Rather interestingly, a
substantial dependence on $c$ of form factors in free space drops out
in the transparency ratio.

\begin{figure} [t,b] \hbox{\hspace{6em}
\hbox{\psfig{figure=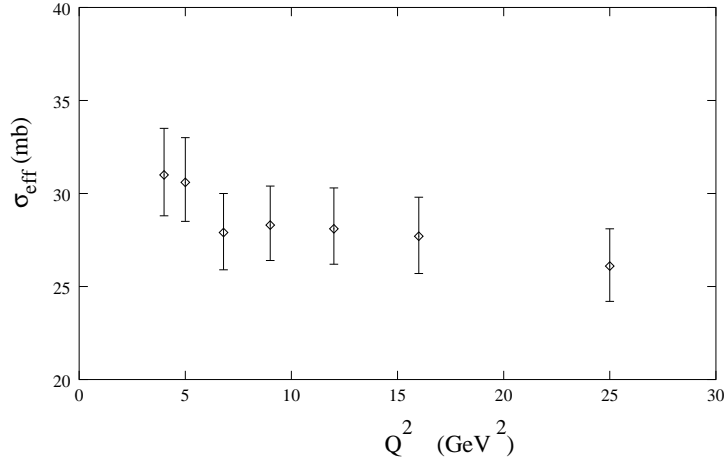,height=6cm}}}
\caption{\em The proton attenuation 
cross section $\sigma_{\rm eff}$ as a function of $Q^2$ extracted
from the monte carlo calculations.
The extracted
value is significantly smaller than the standard Glauber theory value
of 40 mb.}
\label{sigma}
\end{figure} 

If the transparency ratio is flat with $Q^{2}$, does it mean that
color transparency has not been observed?  No, because important
contributions to the {\it free-space denominator} are being filtered away by
a large nucleus, depleting the numerator, making the interpretation of
the $Q^{2}$ dependence ambiguous\cite{JR93}.  Fortunately experiments
also have the $A$ dependence, which can be expressed as an {\it
empirically defined} effective attentuation cross section
$\sigma_{eff}$.  In the standard Glauber theory of $pA$ scattering this
quantity is flat with energy in the region of interest and has a value
of about 40 mb.  Observation of $\sigma_{eff}$ substantially below
the Glauber value would be a signal of color transparency.  We
evaluated our predictions as if they were experimental data, extracted
$\sigma_{eff}$, and found the value to be about 31 mb at $Q^2=4$ GeV$^2$
and slowly decreasing with energy as shown in fig. \ref{sigma}. 
This value is substantially 
below the Glauber
value: so color transparency can be observed in $e,e'p$ measurements
of sufficiently high precision.

Finally in fig.  \ref{pip}b,c we show results for the transparency ratio for
the process $\pi A\rightarrow \pi'p(A-1)$.  Rather interesting
oscillations are predicted for the transparency ratio.  These
predictions can be tested in future experiments.  The results for the
transparency ratio for $\gamma A\rightarrow \pi^+n(A-1)$ are shown in
Fig \ref{gammap}b,c,d.  We again predict fluctuations in the
transparency ratio which can be tested in the near future at CEBAF
\cite{gao104}.

\bigskip
\noindent
{\bf Acknowledgements}: Work was partially supported by DOE grant number 
85ER40214.

\end{document}